\documentclass[aps,pre,superscriptaddress,reprint,showpacs]{revtex4-1}
\pdfoutput=1

\usepackage[pdftex]{graphicx} 
\usepackage{amsmath}
\usepackage{amsfonts}
\usepackage{mathrsfs}
\usepackage{bm}

\usepackage{times}
 
\begin{document}

\title{First-order phase transition and tricritical scaling behavior of the Blume-Capel model: \\ a Wang-Landau sampling approach}

\author{Wooseop Kwak}
\affiliation{Department of Physics, Chosun University, Gwangju 61452, Korea}
\author{Joohyeok Jeong}
\affiliation{Department of Physics and Photon Science, School of Physics and Chemistry, Gwangju Institute of Science and Technology, Gwangju 61005, Korea}
\author{Juhee Lee}
\affiliation{Department of Physics and Photon Science, School of Physics and Chemistry, Gwangju Institute of Science and Technology, Gwangju 61005, Korea}
\author{Dong-Hee Kim}
\email{dongheekim@gist.ac.kr}
\affiliation{Department of Physics and Photon Science, School of Physics and Chemistry, Gwangju Institute of Science and Technology, Gwangju 61005, Korea}

\begin{abstract}
We investigate the tricritical scaling behavior of the two-dimensional spin-$1$ Blume-Capel model by using the Wang-Landau method measuring of the joint density of states for lattice sizes up to $48\times 48$ sites. We find that the specific heat deep in the first-order area of the phase diagram exhibits a double-peak structure of the Schottky-like anomaly appearing with the transition peak. The first-order transition curve is systematically determined by employing the method of field mixing in conjunction with finite-size scaling, showing a significant deviation from the previous data points. At the tricritical point, we characterize the tricritical exponents through finite-size-scaling analysis including the phenomenological finite-size scaling with thermodynamic variables. Our estimation of the tricritical eigenvalue exponents, $y_t = 1.804(5)$, $y_g = 0.80(1)$, and $y_h = 1.925(3)$, provides the first Wang-Landau verification of the conjectured exact values, demonstrating the effectiveness of the density-of-states-based approach in finite-size scaling study of multicritical phenomena.
\end{abstract}


\maketitle

\section{Introduction}

The Wang-Landau (WL) sampling method \cite{WL1,WL2} directly estimates the density of states through random walk in energy space. 
Because of its capability of dealing with complex energy landscape together with the flexibility for applications, it has been widely used in different areas of physics and chemistry, including protein folding \cite{Rathore2002,Wust2009}, fluid simulations \cite{Singh2012}, random spin systems \cite{Okabe2002}, and also quantum systems \cite{Troyer2003,Inglis2013}.
Particularly for study of phase transitions, the WL method suggests an efficient way to overcome the issue of slow dynamics in the conventional Monte Carlo simulations. Reducing tunneling time and critical slowing down in the first- and second-order transitions has been a long-standing subject in the advances of the Monte Carlo methods which include, for instance, the cluster algorithms \cite{Swendsen1987,Wolff1989}, multicanonical ensemble \cite{Berg1992,Zierenberg2015}, parallel tempering  \cite{Swendsen1986,Hukushima1996}, and histogram reweighting technique \cite{Ferrenberg1988}. 
In the WL method, with the density of states being accurately estimated, one can immediately access thermodynamic quantities at any temperatures across phase diagram, indicating its potential for study of critical phenomena (for instance, see \cite{Berg2007,Tsai2007,Vink2010,Silva2006,Hurt2007,Malakis2009,Theodorakis2012}). 
In this paper, we focus on the tricritical phenomena and examine the effectiveness of the Wang-Landau method in the finite-size-scaling analysis of the tricritical behavior in two dimensions.

A tricritical point at which the nature of phase transition changes from first order to second order has been observed in a variety of systems \cite{book:LS}; for instance, multicomponent fluids, metamagnets, $^3$He-$^4$He mixtures~\cite{Bafi2015}, and also recently ultracold quantum gases \cite{Shin2008}.
Interestingly, the upper critical dimension for the Ising tricritical behavior is lowered to three, and thus in two dimensions, the tricritical scaling exponents become different from the classical ones \cite{book:Cardy,book:LandauBinder}. 
The tricritical universality in two dimensions was first conjectured from the dilute Potts model \cite{denNijs1979,Nienhuis1979,Pearson1980} and established by the conformal invariance argument \cite{Nienhuis1982}.
On the other hand, large efforts with advanced numerical methods on different models have been also devoted to precisely calculate the tricritical eigenvalue exponents, namely the thermal exponent $y_t$, the next-to-leading thermal exponent $y_g$, and the magnetic exponent $y_h$.
The Monte Carlo renormalization group (MCRG) calculation was performed for the Ising antiferromagnet (AFM) and the Blume-Capel (BC) model \cite{Landau1981};
the transfer-matrix method was applied to the BC model \cite{Beale1986};
the metropolis algorithm with the histogram reweighting (HR) technique 
was used for the spin fluid and the BC model \cite{Wilding1996}.  
While the WL method was first applied to the BC model in Ref. \cite{Silva2006}, the tricritical scaling exponents still remain unexplored in the same method. 
The tricritical eigenvalue exponents estimated from these previous calculations are listed in Table \ref{table:exponent}.

Here, by using the Wang-Landau method, we approach the tricriticality of the two-dimensional spin-$1$ Blume-Capel model from the side of the first-order phase transitions. The joint density of states is measured for systems with sizes up to $48 \times 48$ sites, allowing an accurate picture of the first-order transitions and tricritical scaling behavior. First, at a large crystal field, we find out a double-peak structure of the specific heat where the Schottky-like anomaly appears together with the first-order transition peak. It turns out that our large-scale calculations are crucial to reveal this anomalous structure. Second, we systematically determine the first-order transition curve which provides significant deviations from the implicit line of the few previous data points. Finally, we characterize the tricritical exponents through finite-size-scaling analysis within the Wang-Landau framework. The excellent data-curve collapses in the finite-size scaling accurately determine the three tricritical eigenvalue exponents, providing the first Wang-Landau verification of the conjectured exact values of the tricritical exponents in two dimensions. 

This paper is organized as follows. 
Section~\ref{sec:method} defines the Hamiltonian of the BC model and provides the numerical details of our simulations. 
In Sec.~\ref{sec:method}, we also briefly describe the mixing-field method with which we proceed for the characterization of the tricritical behavior of the BC model. 
In Sec.~\ref{sec:transition}, we present the specific-heat anomaly observed with the first-order transitions, and we provide the determination of the first-order-transition line with the details of the mixing-field analysis~\cite{Wilding1992,Wilding1996,Plascak2013} that we employ to examine phase coexistence and to locate the tricritical point. 
In Sec.~\ref{sec:fss}, we provide the finite-size-scaling analysis to estimate the tricritical exponents, including the one with the probability distribution associated with the field mixing and the phenomenological finite-size scaling with thermodynamic quantities. 
In Sec.~\ref{sec:conclusions}, summary and outlooks are given.

\begin{table}
\begin{ruledtabular}
\begin{tabular}{ccccc}
Numerical method & Model & $y_t$ & $y_g$ & $y_h$ \\
\hline
MCRG \cite{Landau1981} & Ising AFM, BC & 1.80(2) & 0.84(5) & 1.93(1) \\
Transfer Matrix \cite{Beale1986} & BC & 1.75(3) & 0.80(1) & 1.90(5) \\
Metropolis+HR \cite{Wilding1996} & Spin fluid, BC & 1.80(1) & 0.83(5) & 1.93(1) \\ 
WL (this work) & BC & 1.804(5) & 0.80(1) & 1.925(3) \\ 
\multicolumn{2}{c}{The exact conjectures~\cite{denNijs1979,Nienhuis1979,Pearson1980,Nienhuis1982}} & 9/5 & 4/5 & 77/40 \\
\end{tabular}
\end{ruledtabular}
\caption{Numerical estimations of the tricritical eigenvalue exponents in two dimensions. The conjectured exact exponents are given for comparison.}
\label{table:exponent}
\end{table}

\section{Model and Methods}
\label{sec:method}

\subsection{Grand canonical formulation of the Blume-Capel model}

The spin-1 Blume-Capel model in square lattices that we consider can be described by the Hamiltonian 
\begin{equation}
\label{eq:H}
\mathcal{H} = -J \sum_{\langle i,j \rangle} s_i s_j + \Delta \sum_i s_i^2 - h \sum_i s_i,
\end{equation}
where spin $s_i$ at site $i$ can take a value of $+1$, $-1$, or $0$, and $J$ and $\Delta$ denote the  ferromagnetic coupling and crystal field causing spin anisotropy, respectively.  
The summation $\Sigma_{\langle i,j \rangle}$ runs over all pairs of nearest-neighbor spins. The coupling $J$ is set to be unity to define unit energy scale.
We only consider the case of zero external magnetic field, $h=0$, in the calculations. The system size is denoted by $L$ representing $L^d$ lattice sites where the dimension $d=2$ for our square lattices. 
For the numerical implementation, we write the partition function in a grand canonical form as
\begin{equation}
\label{eq:Z}
\mathcal{Z}_L(\beta,z) = \sum_{E,N} \Gamma(E,N) z^N \exp ( \beta E ),
\end{equation}  
where $\beta$ denotes the inverse temperature $1/k_B T$, and the fugacity $z$ is given as $z \equiv \exp(-\mu)$ with $\mu \equiv \beta\Delta$. 
The Boltzmann constant $k_B$ is set to unity for simplicity. 
The variables $E \equiv \sum_{\langle i,j \rangle} s_i s_j$ and $N \equiv \sum_i s_i^2$ represent the kinetic energy and number of nonzero spins, respectively. 
The joint density of states $\Gamma(E,N)$ is to be given by the WL sampling.  

\subsection{Direction of scaling fields}

We employ the method of field mixing~\cite{Wilding1992,Wilding1996,Plascak2013} to describe the asymmetry of phase transition undergoing in the Blume-Capel model and the scale invariance at the tricritical point. 
The formulation in Eqs. (\ref{eq:H}) and (\ref{eq:Z}) suggests the temperature $T$, crystal field $\Delta$, and magnetic field $h$ as a natural choice of fields to describe the phase diagram.
While the scaling direction associated with $h$ is orthogonal to the $T-\Delta$ (or $\beta-\mu$) plane because of the Ising symmetry, there is no such symmetry for the other two fields. 
Thus, for instance in order to study the tricritical behavior, one may consider the linear combinations of $\beta$ and $\mu$ to describe the relevant scaling fields as
\begin{eqnarray}
\label{eq:scalingfield}
\lambda &=& (\mu-\mu_t) + r (\beta-\beta_t), \label{def:l} \\
g &=& (\beta-\beta_t) + s (\mu-\mu_t), \label{def:g} 
\end{eqnarray}
where $\mu_t$ and $\beta_t$ are the values at the tricritical point, and $r$ and $s$ are the mixing parameters. 
The scaling field $g$ is tangent to the coexistent curve while the direction of $\lambda$ is not restricted. 
Accordingly, one can also write down the two relevant variables conjugate to the scaling fields as
\begin{eqnarray}
\mathcal{Q}&=&\frac{1}{1-rs}(n-s\epsilon), \label{def:Q} \\
\mathcal{E}&=&\frac{1}{1-rs}(\epsilon-rn), \label{def:E}
\end{eqnarray}
where $n=L^{-d}N$ and $\epsilon=L^{-d}E$, satisfying the requirement $\langle X \rangle = L^{-d}\partial \ln \mathcal{Z}_L / \partial x$ for scaling field $x$.
Note that the mixing parameters $r$ and $s$ are system-specific quantities, and thus the scaling fields and their corresponding conjugate variables can exhibit system-size dependence.  

In the vicinity of the tricritical point, the finite-size-scaling ansatz for the limiting probability  distribution function of the scaling fields and their conjugate variables is written as
\begin{eqnarray}
\label{eq:P}
P_L & \propto \tilde{p}_L ( & a_t^{-1} L^{d-y_t} \mathcal{Q}, 
a_g^{-1} L^{d-y_g} \mathcal{E}, a_h^{-1} L^{d-y_h} m, \nonumber \\
& & a_t L^{y_t} \lambda, a_g L^{y_g} g, a_h L^{y_h} h),
\end{eqnarray}
where $\tilde{p}_L$ is a universal scaling function, and $a$'s are nonuniversal factors (for more details, see Refs. \cite{Wilding1992,Wilding1996,Plascak2013}).  
Precisely at the tricritical point, the probability distribution function becomes
\begin{equation}
P_L \propto \tilde{p}^*_L (a_t^{-1} L^{d-y_t} \mathcal{Q}, 
a_g^{-1} L^{d-y_g} \mathcal{E}, a_h^{-1} L^{d-y_h} m),
\label{eq:P*}
\end{equation}
where $\tilde{p}^*_L$ is universal and scale invariant, which allows use to measure the tricritical 
exponents $y$'s from the finite-size scaling for systems with different sizes.
The probability distribution function of the field-conjugate variables can be estimated from the histogram accumulating the occurrence of $(E,N)$-points in the discrete bins of $\mathcal{Q}$ and $\mathcal{E}$ with weighting factor $\Gamma(E,N) z^N \exp(\beta E)$.   
 
\subsection{Numerical aspects of the Wang-Landau sampling}

Our numerical estimation of $\Gamma(E,N)$ follows the standard WL algorithm~\cite{WL1,WL2} except that our random walk needs to be performed in the two dimensional parameter space of $E$ and $N$. Initially, the density of states $\Gamma(E,N)$ is set to be unity, and the random walk proceeds by trying out a new random value for a spin randomly chosen in the lattices. The new trial spin would move the energy from $(E_1,N_1)$ to $(E_2,N_2$) in the two dimensional parameter space, and then this spin update is accepted with the transition probability
\begin{equation}
p[(E_1,N_1)\to (E_2,N_2)] = \min \left( \frac{\Gamma(E_1,N_1)}{\Gamma(E_2,N_2)}, 1 \right)
\end{equation}
for the sake of the importance sampling of $\Gamma$. In every trial of spin updates, the current energy state $(E,N)$ is recorded in $\Gamma$ and the histogram $H$ of visited states as
$\ln\Gamma(E,N) \to \ln\Gamma(E,N) + \ln f$ with the modification factor $f$ and $H(E,N) \to H(E,N) + 1$, respectively. These WL procedures continues until the histogram becomes flat enough and then are restarted  with a reduced modification factor and with resetting the histogram as $H=0$. 

The modification factor $f$ is initially given as $\ln f=1$ and scaled down as $\sqrt{f}$ when restarting.
The flatness criterion for the histogram is set to be $95 \%$ for $L=16$ and $90 \%$ for $L=20$ and $24$, and it is lowered to $80 \%$ for $L\ge 32$. 
To avoid accidental satisfaction of the flatness criterion, the number of Monte Carlo steps (MCS) between successive flatness inquiries is set to be the same as the number of available energy states of $(E,N)$, where unit MCS is defined as $L^d$ trials of a single spin update. 
The actual flatness inquiry interval is about $10^5$ MCSs for $L=16$ and increases to $10^7$ for $L=48$. 
The stopping criterion for the modification factor is given as $\ln f< 10^{-8}$ for $L \le 32$, a less stringent $10^{-7}$ for $L=40$, and $10^{-6}$ for $L=48$. 

The main difficulty encountered in these procedures comes from the increased dimensionality encoded in the importance sampling with spin update trial, causing very long computation times. Compared to the usual case with a single energy parameter, many more spin update trials are needed to cover the two-dimensional space of $(E,N)$ since one spin update trial can visit only one energy state. For $L=48$, the size of $(E,N)$-space is enlarged to be about $10^7$, while the corresponding number for the Ising energy $E$ is just in the order of thousands.   
Therefore, the WL simulations for multi-energy variables cost significantly more in computational time than the one-variable case does (see also Refs. \cite{Silva2006,Zhou2006,Tsai2007}). 
For instance, our computation of $\Gamma(E,N)$ for the system with $L=40$ takes about six months on a 3.3 GHz Xeon E3 processor. While the WL procedures that we consider here is standard and essentially serial, extending the recently suggested broad kernel update method \cite{Zhou2006,Junghans2014} and massively parallel algorithm ~\cite{Vogel2013,Vogel2014} to a multiparameter system may help to reduce the issue of long computation time. 

Once $\Gamma(E,N)$ is obtained from the WL procedures, it is straightforward to calculate the canonical average of a thermodynamic observable $\mathcal{O}\equiv \mathcal{O}(E,N)$  at given $T$ and $\Delta$ as 
\begin{equation}
\label{eq:Oavg}
\langle \mathcal{O} \rangle \equiv  \frac{1}{\mathcal{Z}}\sum_{E, N} \mathcal{O}(E,N) \Gamma(E,N) z^N \exp(\beta E).
\end{equation}
Similarly, one can also define the moment of microcanonical magnetization as
\begin{equation}
\label{eq:mavg}
\langle |m|^k \rangle \equiv  \frac{1}{\mathcal{Z}}\sum_{E, N} [\langle|m|\rangle_{E,N}]^k \Gamma(E,N) z^N \exp(\beta E),
\end{equation}
where the microcanonical magnetization $\langle |m|\rangle_{E,N}$ is an average of $|m| \equiv L^{-d}|\sum_i s_i|$ for a given $(E,N)$ which can be measured simultaneously with the WL sampling~\cite{Silva2006,Caparica2012}. 
In practice, the microcanonical average is performed in the last stage of the iterations with the smallest $f$ where the density of states is saturated. 
In our simulations, the random walk done for convergence of the microcanonical magnetization is typically in the order of a thousand flatness inquiries; however, we find that the estimation of $\langle |m|\rangle_{E,N}$ is still numerically affordable for $L\le 40$ within the limited computational time. 
We calculate the susceptibility and fourth-order cumulant of microcanonical magnetization by using Eq. (\ref{eq:mavg}). 
While the moment of microcanonical magnetization may quantitatively differ from the genuine canonical counterpart, our finite-size-scaling analysis given in the later sections shows that it is still very useful for the estimate of the first-order-transition points and, more importantly, it shares the same universal behavior anticipated at the tricritical point.     

\section{Phase diagram of the Blume-Capel model}
\label{sec:transition}

\begin{figure}
\includegraphics[width=0.48\textwidth]{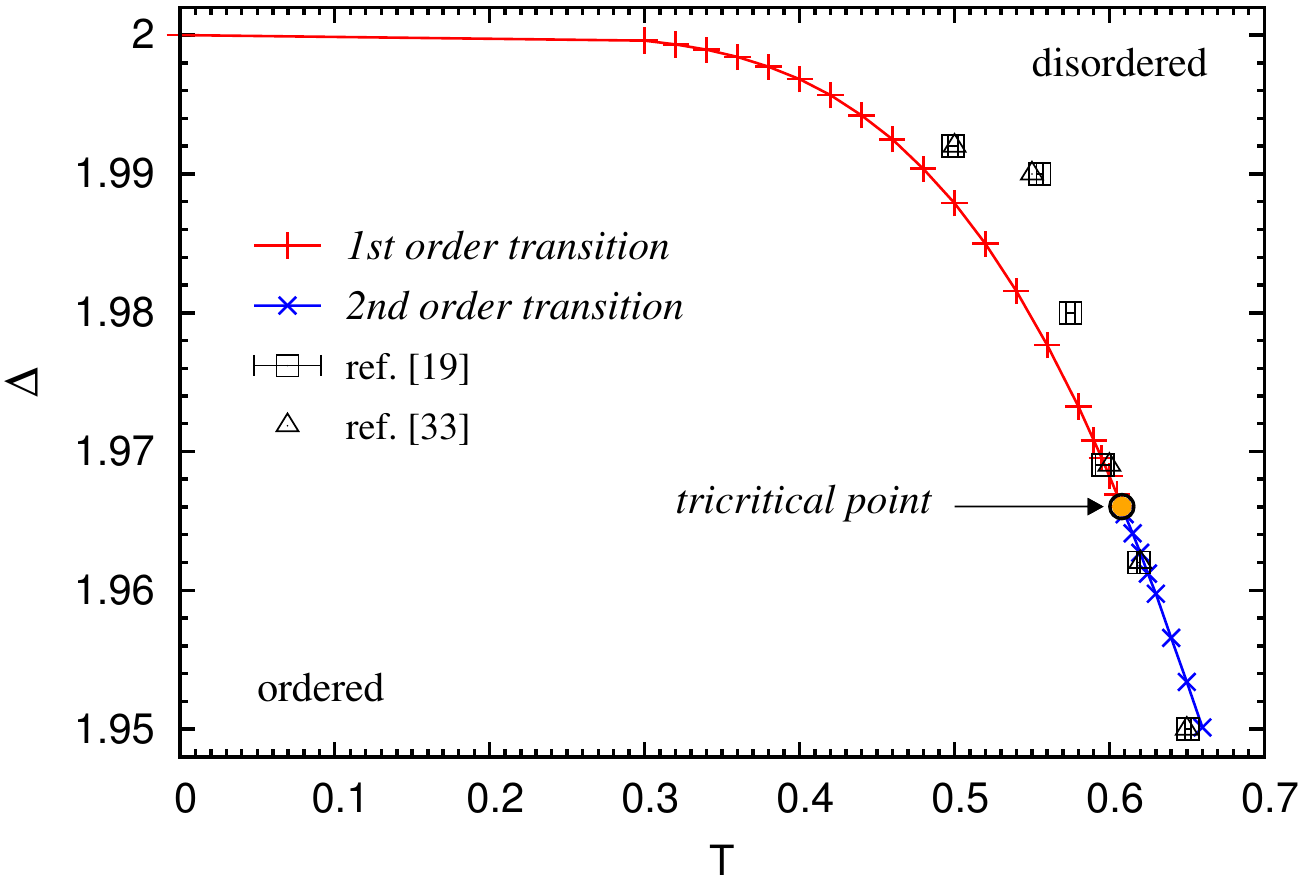}
\caption{\label{fig1}
(Color online) Phase diagram of the two-dimensional spin-1 Blume-Capel model around the tricritical point in the plane of temperature $T$ and crystal field $\Delta$. The transition line and tricritical point are determined from the mixing-field analysis with the calculations using the Wang-Landau density of states. The phase-transition line plotted here is determined in the limit of infinite $L$ from the extrapolation of the pseudotransition points obtained for the systems with different sizes up to $L=48$ (for example, see Fig. \ref{fig3}). The transition points previously available from the literature \cite{Beale1986,Silva2006,footnote_data} are given for comparison. The statistical error was unspecified for the data in Ref. [33], and the error bars of the data in Ref. [19] were given in temperature. All our transition points are listed in Table~\ref{table:firstorder}.
}
\end{figure} 

In this section, we particularly focus on the area of the phase diagram of the BC model at large crystal fields just below $\Delta=2$. It is known that the first-order transitions dominates in this area, however, the detailed plot of the first-order transition line is not available yet. 
In Fig.~\ref{fig1}, we present the transition line that we obtain as a function of temperature and crystal field, where the estimated location of the tricritical point is also specified. 
While our estimation of the tricritical point is in good agreement with the previous numerical results~\cite{Beale1986,Wilding1996,Silva2006,Plascak2013,Xavier1998}, the transition points available from the literature~\cite{Beale1986,Silva2006} shows a significant deviation from the first-order transition line that we identify with the WL method. The finite-size scaling for the extrapolation is performed with system sizes $L\le 48$, and these large-scale calculations are also crucial to reveal the valid physics of the specific heat occurring deep in the first-order area.

\begin{figure}
\includegraphics[width=0.45\textwidth]{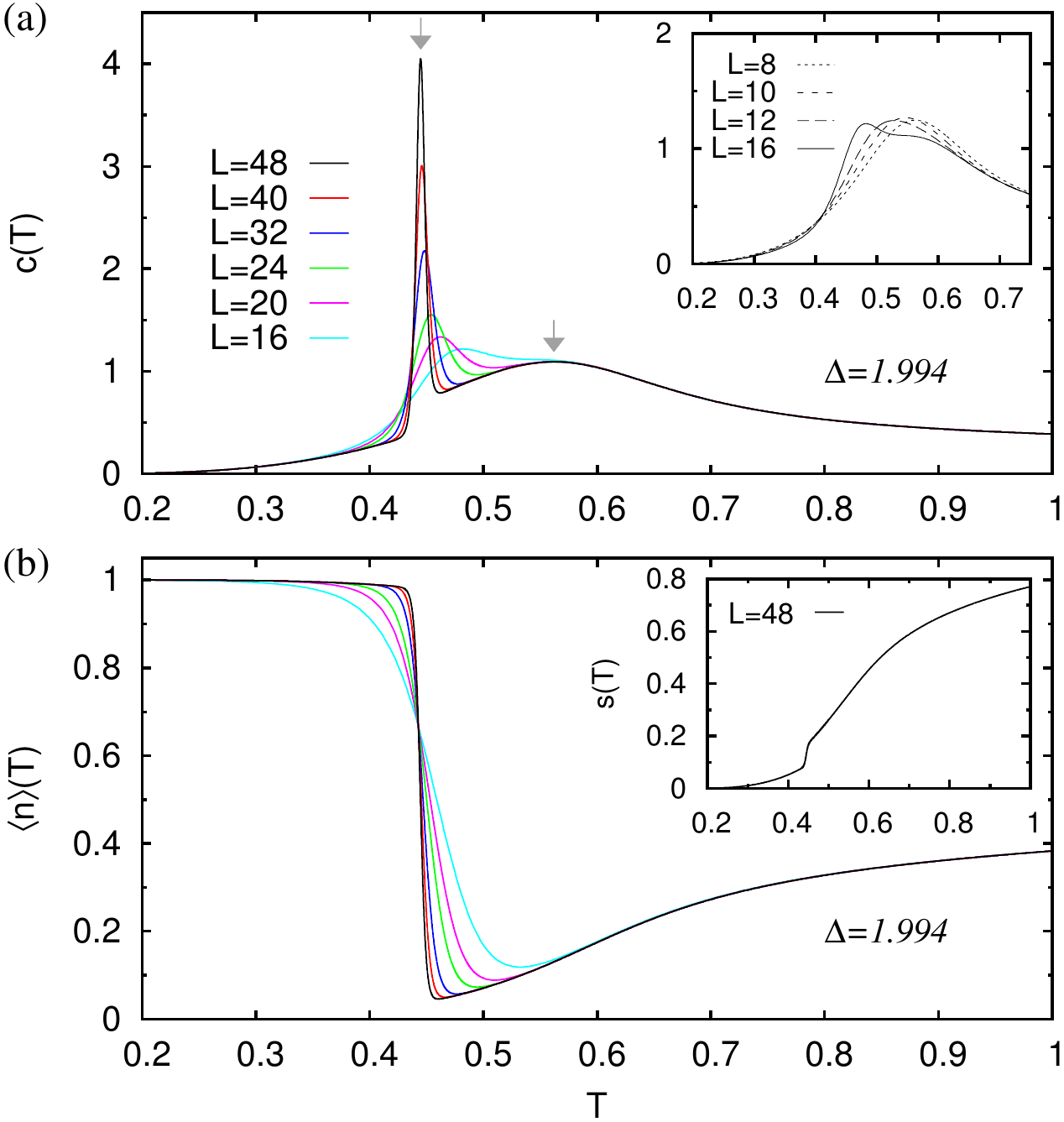}
\caption{\label{fig2}
(Color online) Specific-heat anomaly appearing with the first-order transition. (a) Double-peak structure observed in the specific heat at $\Delta=1.994$. While the diverging peak at the lower temperature is associated with the phase transition, the anomalous broad one at the higher temperature does not scale with the system size. The inset indicates that the smaller systems cannot detect the correct structure. (b) Corresponding density of nonzero spins $\langle n \rangle$ and entropy per site $s(T)$. The zero-spin state is dominant right after the transition and then starts to get thermally excited to the nonzero-spin states distributed with increasing entropy.
}
\end{figure}

Our Wang-Landau calculations also reveal an anomalous double-peak structure of the specific heat at large crystal field, yet in the first-order-transition area. Figure~\ref{fig2} displays the structure of the specific heat at $\Delta=1.994$ where the broad anomaly emerges above the sharp divergence of the first-order transition. This anomaly does not scale with system size and is associated with the Schottky-like mechanism. We find that, at the first-order transition, the population of nonzero spins sharply drops on the disordered side. Since the zero spins dominate at this stage, similar to the Schottky anomaly, the energy barrier for the excitation of the nonzero-spin states is mainly from the crystal field $\Delta$. While this anomaly could be anticipated for very strong anisotropy with $\Delta/J \gg 1$, it was not known whether it could appear together with the phase transition. The previous WL calculations for the specific heat are limited to $L\le 16$~\cite{Silva2006} which we find cannot detect the double-peak structure [see the inset of Fig.~\ref{fig2}(a)]. In our calculation, for system sizes up to $L=48$, the double-peak structure of the specific heat is visible for $\Delta \ge 1.99$.      

\subsection{First-order phase transitions}

\begin{figure}
\includegraphics[width=0.48\textwidth]{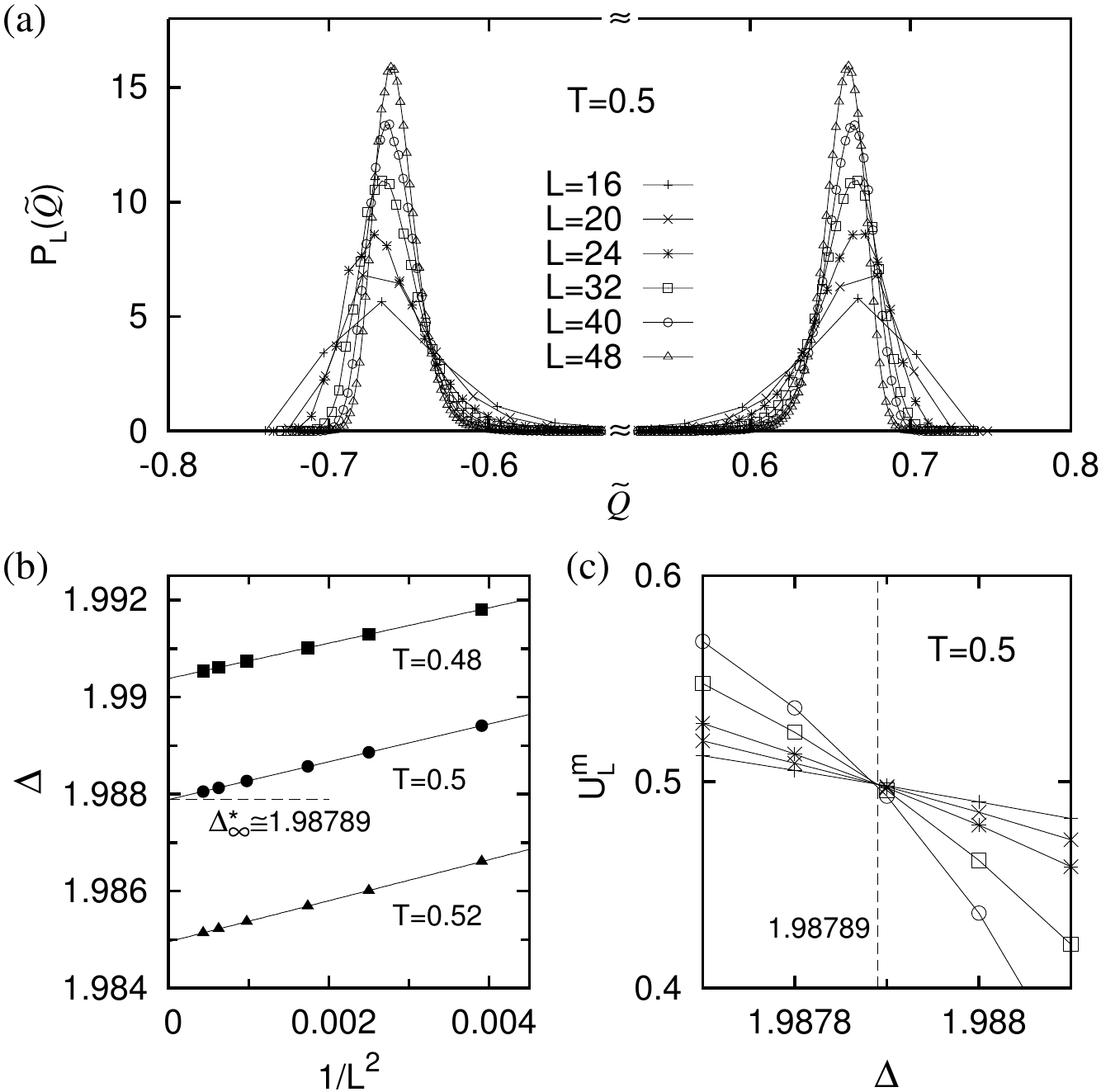}
\caption{\label{fig3}
Finding the first-order transition points by using the method of field mixing. (a) Symmetric double peaks in the distribution of the field-conjugate variable $\mathcal{Q}$ at the phase coexistence $\Delta=\Delta^*_L$ established at temperature $T=0.5$. The distribution is shifted and rescaled for visualization of systems with different sizes. (b) Scaling behavior of $\Delta^*_L$ with system size $L$. The extrapolation in the limit of $L \to \infty$ determines the transition point $\Delta^*_\infty$. The fourth-order cumulant of microcanonical magnetization $U^m_L$ in panel (c) also find its crossing point very close to the estimated transition point.
}
\end{figure}   

We determine the first-order phase transition line from the symmetry condition for phase coexistence in the probability distribution function $P_L(\mathcal{Q})$. 
The energy variable $\mathcal{Q}$ is conjugate to the scaling field $\lambda$ across the phase transition, and thus one may expect a symmetric doubly peaked $P_L(\mathcal{Q})$ at the transition point in analogy with the probability distribution of the order parameter in conventional first-order transitions \cite{Wilding1996}. 
For our systems with finite sizes, we search for a size-dependent pseudotransition point and mixing parameter at which the symmetric doubly peaked $P_L(\mathcal{Q})$ emerges.  

Figure~\ref{fig3}(a) illustrates the symmetric probability distribution with the double peaks found for the phase coexistence at a given temperature. 
In the graphical search for the symmetry to find the pseudotransition point $\Delta^*_L$ and mixing parameter $s$, a practical difficulty lies in discriminating the shape of the distribution, which actually is the histogram of the discrete values of $\mathcal{Q}$ constructed with finite bin size.
For the optimized identification of the symmetry, we compare various local statistics of the double peaks, including populations and heights. 

\begin{figure}
\includegraphics[width=0.375\textwidth]{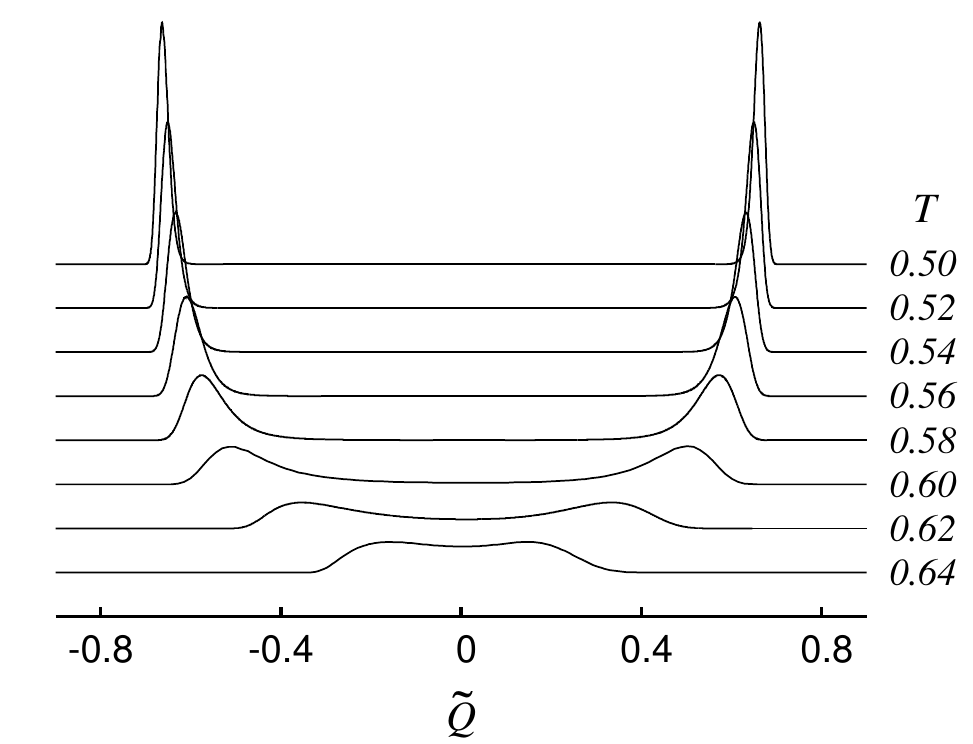}
\caption{\label{fig4}
Probability distribution function of the field-conjugate variable $\mathcal{Q}$ along the transition line. The calculations for the largest available system with $L=48$ are shown.
}
\end{figure}

The search occurs in three main steps. 
First, for a given $T$, we graphically search for a set of $\Delta$ and $s$ that roughly gives double peaks in $P_L(\tilde{\mathcal{Q}})$ where $\mathcal{Q}$ is normalized for $\tilde{\mathcal{Q}}$ to have zero average and unit variance. 
Second, starting from these initial values, $\Delta$ is fine-tuned to meet the equal population condition by measuring the difference in population below and above $\tilde{\mathcal{Q}}=0$. 
In this step, we also check the symmetry of the local averages measured for the parts below and above the zero point, which we find comes along with the equal population condition. 
Note that this step is independent of any graphical visualization and therefore allows the high-resolution determination of $\Delta^*_L$ in the WL approach. 
Then, with $\Delta$ being fixed, the mixing parameter $s$ can be determined graphically for the condition of equal height of the double peaks. 
We find that the two peaks are well separated in the first-order transition region as explicitly shown in Fig.~\ref{fig4}, and the tuning of $s$ mainly changes the peak heights without disturbing the equal population condition. 
In our numerical implementation, we determine the pseudotransition point $\Delta^*_L$ when the difference in population and height is minimized within the search step of $10^{-6}$ in $\Delta$ for $T<0.6$ and $10^{-5}$ otherwise to get enough resolution for size scaling. 

We obtain the transition point $\Delta^*_\infty$ from the extrapolation of the pseudotransition point $\Delta^*_L$ in the limit of $L \to \infty$. 
For the area of the phase diagram shown in Fig.~\ref{fig1}, we find that $\Delta^*_L$ shows the scaling behavior
\begin{equation}
\label{eq:Dinf}
\Delta^*_L  = \Delta^*_\infty + b L^{-2} ,
\end{equation}
where $b$ is a fitting parameter, which agrees well with the $L^{-d}$ scaling generally expected for the first-order phase transitions. 
The scaling behavior around $T=0.5$ is presented in Fig.~\ref{fig3}(b), for example. 
We also examine the crossing point of the fourth-order cumulant of the microcanonical magnetization $U^m_L \equiv 1-\langle |m|^4 \rangle / 3\langle |m|^2 \rangle^2$ measured for systems with different sizes. 
We find that the crossing of $U^m_L$ is in good agreement with the transition point $\Delta^*_\infty$ obtained from the analysis of the probability distribution [for instance, see Fig.~\ref{fig3}(c)]. 
The difference between the two different approaches is observed to be about $10^{-4}$ for $0.58<T<0.64$ beyond the estimated errors which could be further improved by averaging over many samples of the WL density of states. 
The estimated transition points are listed in Table.~\ref{table:firstorder}.

\begin{table}[t]
\begin{ruledtabular}
\begin{tabular}{lccc}
$T$ & $\Delta^*_\infty$ & crossing of $U^m_L$ & order of transition \\
\hline
0.3   & 1.99960(1) & 1.99960 & first \\
0.32  & 1.99933(1) & 1.99932 & first \\
0.34  & 1.99895(1) & 1.99894 & first \\
0.36  & 1.99842(1) & 1.99842 & first \\
0.38  & 1.99772(1) & 1.99772 & first \\
0.40  & 1.99681(1) & 1.99681 & first \\
0.42  & 1.99566(1) & 1.99566 & first \\
0.44  & 1.99423(1) & 1.99423 & first \\
0.46  & 1.99248(1) & 1.99248 & first \\
0.48  & 1.99038(1) & 1.99038 & first \\
0.5   & 1.98789(1) & 1.98789 & first \\
0.52  & 1.98496(1) & 1.98496 & first \\
0.54  & 1.98157(1) & 1.98157 & first \\
0.56  & 1.97766(1) & 1.97766 & first \\
0.58  & 1.97323(1) & 1.97322 & first \\
0.59  & 1.97080(1) & 1.97077 & first \\
0.595 & 1.96953(1) & 1.96949 & first \\
0.6   & 1.96825(1) & 1.96817 & first \\ 
0.605 & 1.96690(1) & 1.96681 & first \\
0.608 & 1.96604(1) & 1.96597 & tricritical point \\
0.61  & 1.96550(1) & 1.96541 & second \\
0.615 & 1.96412(1) & 1.96399 & second \\
0.62  & 1.96270(1) & 1.96253 & second \\
0.625 & 1.96125(2) & 1.96106 & second \\
0.63  & 1.95980(5) & 1.95954 & second \\
0.64  & 1.9565(1)  & 1.95647 & second \\
0.65  & 1.9534(1)  & 1.95331 & second \\
0.66  & 1.9501(1)  & 1.95006 & second \\ 
\end{tabular}
\end{ruledtabular}
\caption{Estimated transition points. The extrapolated values of $\Delta^*_\infty$ obtained from the size scaling in Eq.~(\ref{eq:Dinf}) and the crossing points of the fourth-order cumulant $U^m_L$ are given.}
\label{table:firstorder}
\end{table}

The scaling behavior in Eq.~(\ref{eq:Dinf}) certainly supports the first-order characteristics of the transition occurring in the area of the phase diagram that we are after. 
Within our data obtained for systems with sizes up to $L=48$, we have not found a quantifiable change of the scaling behavior which, on the other hand, one may expect to see above the tricritical point of the BC model where the second-order transition should emerge. However, in the probability distribution shown in Fig.~\ref{fig4}, we find that the positions of the double peaks in $P_L(\tilde{\mathcal{Q}})$ get closer as the temperature increases. Above $T=0.64$, the peaks start to merge together in the larger systems, which implies that the character of the transition indeed alters.  

\subsection{Tricritical Point}
\label{sec:tp}

We determine the precise location of the tricritical point from the scale-invariant universal form of the probability distribution function $P_L(\mathcal{Q})$, as indicated in Eq. (\ref{eq:P*}). 
The scale invariance at the tricritical point can be conveniently indicated by a size-independent crossing point of the fourth-order cumulant 
\begin{equation}
U^\mathcal{Q}_L \equiv 1-\frac{\langle {\tilde{\mathcal{Q}}}^4 \rangle}{3\langle \tilde{\mathcal{Q}}^2 \rangle^2}
\end{equation} 
for the field-conjugate variable $\tilde{\mathcal{Q}}$ normalized to have zero average and unit variance \cite{Wilding1996,Silva2006,Plascak2013}.

\begin{figure}
\includegraphics[width=0.48\textwidth]{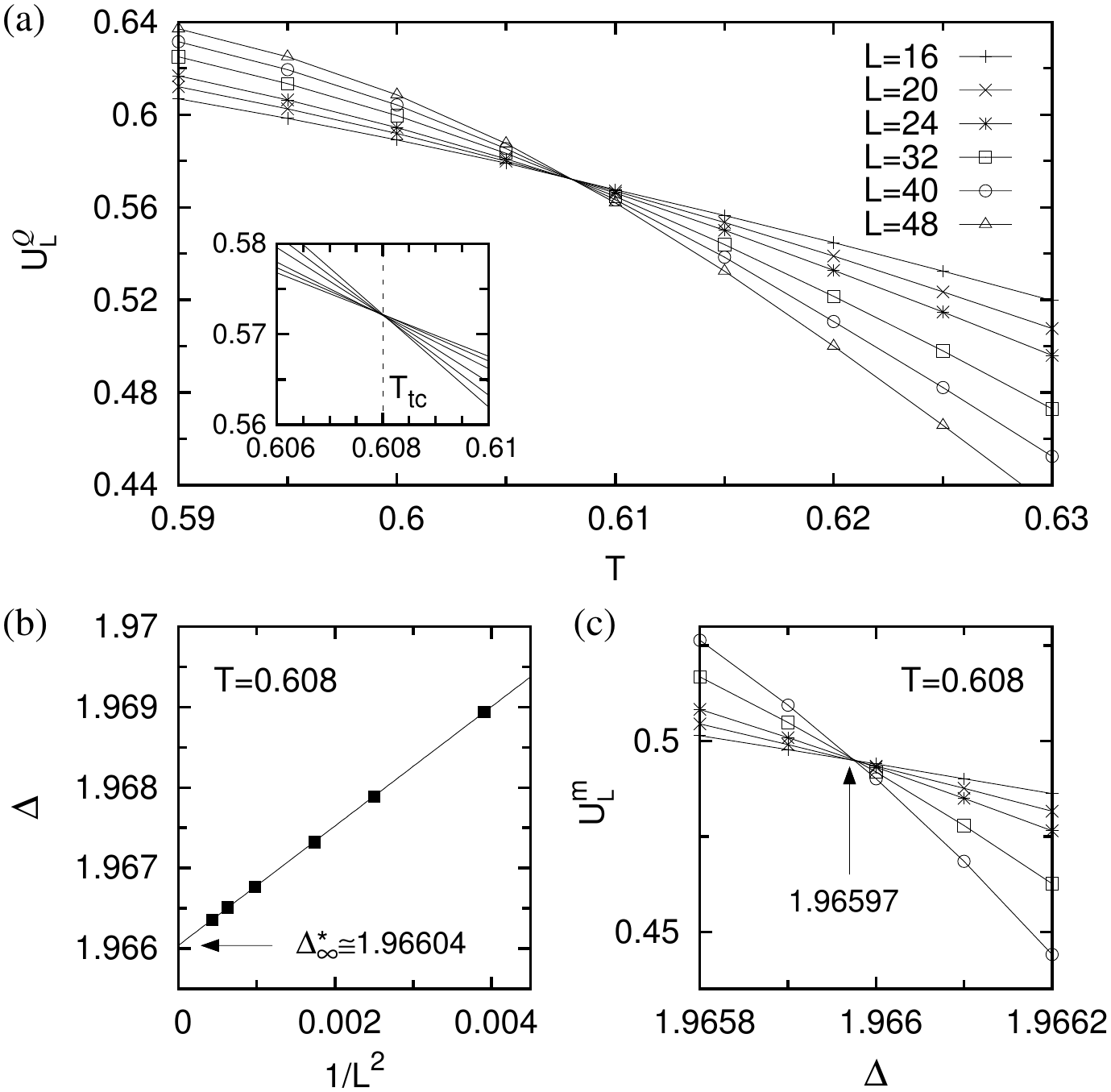}
\caption{\label{fig5}
Location of the tricritical point. (a) The tricritical temperature $T_{tc}\simeq 0.6080$ is determined at the crossing point of the fourth-order cumulant of the field-conjugate variable $U^{\mathcal{Q}}_L$ along the transition line $\Delta=\Delta^*_L(T)$. The extrapolation of the transition points $\Delta_L^*(T_{tc})$ in (b) and the crossing point of the fourth-order cumulant of microcanonical magnetization $U^m_L$ in (c) provide the estimation of the tricritical crystal field as $\Delta_{tc}\simeq 1.9660(1)$. 
}
\end{figure}

We identify the tricritical temperature as $T_{tc} = 0.6080(1)$ from the location of the crossing point of the fourth-order cumulant $U^\mathcal{Q}_L$ along the transition line as shown in Fig.~\ref{fig5}(a). 
Note that the transition line here is for finite $L$; namely, the line of the pseudotransition points $\Delta^*_L(T)$ that we have determined for the phase coexistence. 
The error estimation is only graphical since the our calculation is based on a single sample of the WL density of states. 
We estimate the tricritical crystal field as $\Delta_{tc}=1.9660(1)$ from the extrapolation of the pseudotransition point $\Delta_L^*$ and also from the crossing point of the fourth-order cumulant  $U^m_L$ measured at $T=T_{tc}$ [see Figs. \ref{fig5}(b) and \ref{fig5}(c)]. 
Our estimate of the tricritical point, $T_{tc} = 0.6080(1)$ and $\Delta_{tc} = 1.9660(1)$, is in very good agreement with the previous results for the spin-$1$ Blume-Capel model in square lattices, which provide $T_{tc}=0.610(5)$ and $\Delta_{tc}=1.965(5)$ \cite{Beale1986}, $T_{tc}=0.608(1)$ and $\Delta_{tc}=1.9665(3)$ \cite{Wilding1996}, $T_{tc}=0.609(4)$ and $\Delta_{tc}=1.965(5)$~\cite{Xavier1998}, $T_{tc}=0.609(3)$ and $\Delta_{tc}=1.966(2)$ \cite{Silva2006}, and very recently $T_{tc}=0.608(1)$ and $\Delta_{tc}=1.9665(3)$ \cite{Plascak2013}.

\section{Tricritical Scaling Behavior}
\label{sec:fss}

In this section, we present the three different forms of finite-size-scaling analysis that we perform to determine the tricritical eigenvalue exponents. 
The thermal exponent $y_t$ is extracted from the probability distribution function of the field-conjugate variable $\mathcal{Q}$ at the tricritical point. 
The scaling of the fourth-order cumulant $U^{\mathcal{Q}}_L$ along the transition line is examined to obtain the next-to-leading thermal exponent $y_g$. 
Finally, we perform the phenomenological finite-size-scaling analysis with thermodynamic quantities including specific heat, compressibility, susceptibility, magnetization to measure the thermal and magnetic exponents $y_t$ and $y_h$.

\subsection{Distribution of the field-conjugate variable}

We examine the tricritical thermal exponent $y_t$ from the probability distribution function given in Eq. (\ref{eq:P*}). 
Precisely at the tricritical point, $T=T_{tc}$ and $\Delta=\Delta^*_L(T_{tc})$, the distribution function for the relevant field-conjugate variable $\tilde{\mathcal{Q}}$ can be reduced into the simple finite-size-scaling ansatz \cite{Wilding1996} as
\begin{equation}
\label{eq:fss_p}
P_L(\tilde{\mathcal{Q}}) = L^{d-y_t} p^*_{\mathcal{Q}}(L^{d-y_t}\tilde{\mathcal{Q}}),
\end{equation}
where $p^*_{\mathcal{Q}}$ is a universal function and the dimension is given as $d=2$ for square lattices. 

\begin{figure}
\includegraphics[width=0.48\textwidth]{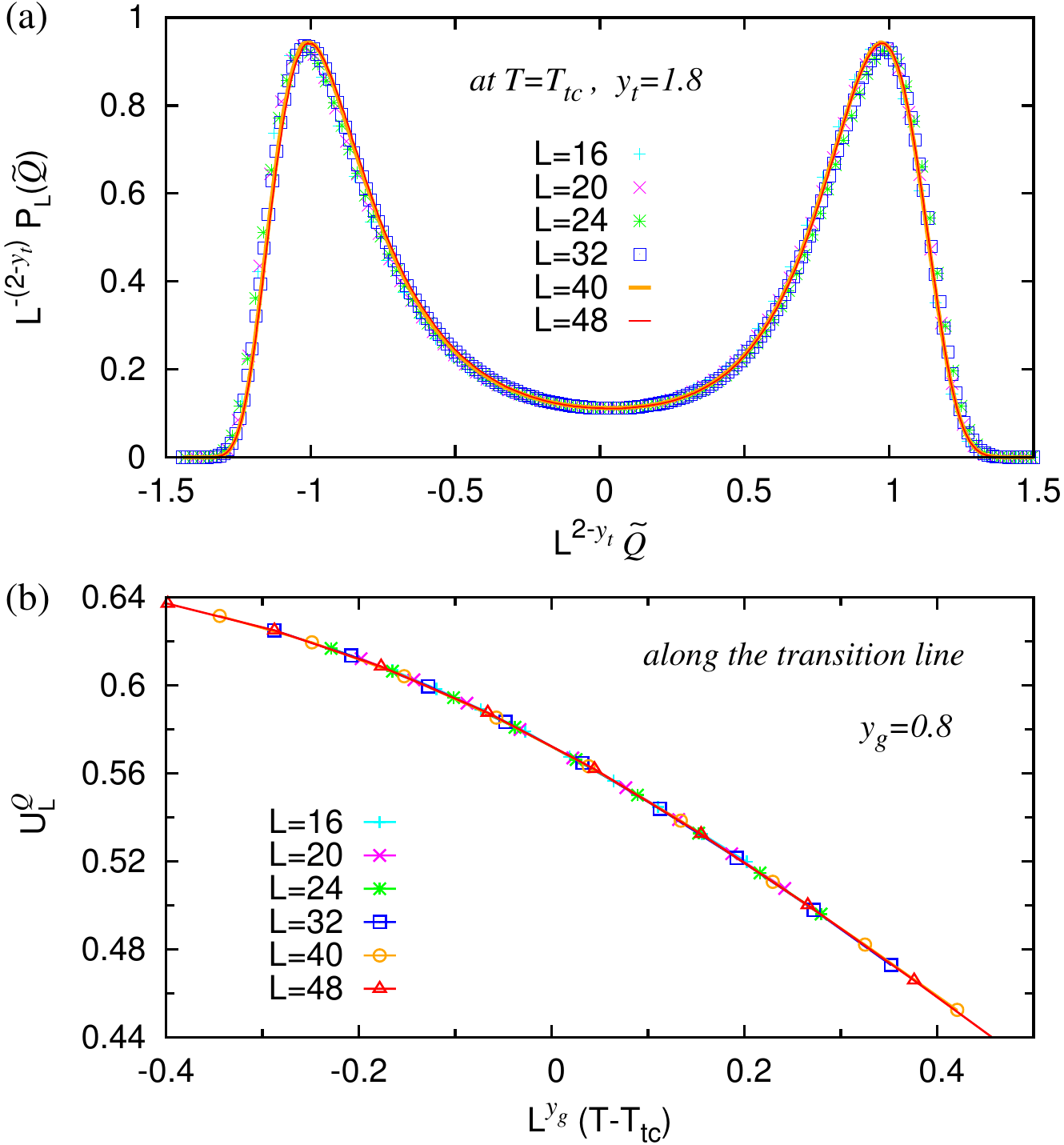}
\caption{\label{fig6}
(Color online) Finite-size-scaling tests of the field-conjugate variable $\mathcal{Q}$ for the tricritical thermal exponents. (a) Scaling plots of the probability density distribution $P_L(\tilde{\mathcal{Q}})$ with the exponent $y_t=1.8$ at the tricritical temperature $T_{tc}=0.608$. The large systems with $L\ge 40$ provide smooth curves (solid lines) falling on a universal distribution which also fits well with the data points for the smaller systems (symbols). (b) Finite-size scaling of the fourth-order cumulant $U^{\mathcal{Q}}_L$ along the transition line. The data points fall onto each other very well, given the next-to-leading exponent $y_g = 0.8$.   
}
\end{figure}

Figure~\ref{fig6}(a) presents our finite-size-scaling analysis for the probability distribution with the tricritical thermal exponent $y_t=1.80(1)$, showing the data of $P_L(\tilde{\mathcal{Q}})$ falling well onto a single curve. 
In particular, the lines for $L=40$ and $48$ can hardly be distinguished in the plot because of the almost perfect overlap. 
The possible error in this estimate with the shape of the distribution mainly originates from the discrete nature of $\mathcal{Q}$ which affects the visualization of its histogram, particularly in small systems, and thus can cause ambiguity in the graphical determination of the mixing parameter. 

Our estimation of $y_t\simeq 1.80$ numerically confirms the exact conjecture $y_t=9/5$ within the graphical identification. The data collapse of $P_L(\tilde{\mathcal{Q}})$ for systems with different sizes $L=16$ to $48$ shows good agreement with the previous finite-size scaling for the spin fluid model, which was also compared for universality with the BC model with size $L=40$~\cite{Wilding1996}. 
In principle, one can also attempt to extract the next-to-leading exponent $y_g$ from the similar finite-size scaling of the probability distribution function $P_L(\mathcal{E})$ as implied in Eq.~(\ref{eq:P*}). 
However, we find that $P_L(\mathcal{E})$ does not give any meaningful estimate of $y_g$ because the distribution is too close to the Gaussian normal distribution, regardless of the system size $L$. 
The same issue was also reported by the previous work~\cite{Wilding1996} where $y_g=1.03(7)$ was estimated from the finite-size-scaling test of $P_L(\mathcal{E})$. 

\subsection{Fourth-order cumulant along the transition line}

Instead, we utilize the fourth-order cumulant $U^\mathcal{Q}_L$ for the estimate of the next-to-leading thermal exponent $y_g$. 
From Eq. (\ref{eq:P}) as well as from the scaling hypothesis of the persistence length~\cite{Beale1986}, one can find that the finite-size scaling of $U^\mathcal{Q}_L$ \textit{along the transition line} may follow the scaling form $U^\mathcal{Q}_L = u^*[L^{y_g}g]$, where $u^*$ is a universal function, and the scaling field $g$ is the deviation from the tricritical point in the direction tangent to the coexistence curve. 
Moreover, in our observation of the data for the phase diagram, it turns out that $(\mu-\mu_t)$ is almost linearly proportional to $(\beta-\beta_t)$ along the transition line near the tricritical point, which leads to $g \sim (T-T_{tc})$ in Eq.~\ref{def:g}. Therefore, for the explicit finite-size scaling tests, one can further simplify the scaling ansatz of $U^\mathcal{Q}_L$ as 
\begin{equation}
\label{eq:fss_u}
U^\mathcal{Q}_L |_{\Delta=\Delta^*_L(T)} \approx u^*[L^{y_g}(T-T_{tc})],
\end{equation}
where the constraint $\Delta=\Delta^*_L(T)$ ensures that it is along the transition line for a system with finite size $L$.

Figure~\ref{fig6}(b) shows that our data points of $U^\mathcal{Q}_L$ along the transition line fall perfectly onto the same curve in the test with $y_g=0.8$ for Eq.~(\ref{eq:fss_u}). 
Within the graphical uncertainty, we determine the next-to-leading thermal exponent $y_g=0.80(1)$, which confirms the exact conjecture $y_g=4/5$. 
Our finite-size-scaling analysis of $U^\mathcal{Q}_L$ can be compared with the finite-size-scaling test of the persistence length, which indicates $y_g=0.80(1)$~\cite{Beale1986} and the estimate made by using the slope of the fourth-order cumulant which provides $y_g=0.83(5)$~\cite{Wilding1996}.    

\subsection{Phenomenological finite-size scaling} 

In this section, we present the phenomenological finite-size-scaling analysis of thermodynamic quantities to determine the thermal and magnetic exponents $y_t$ and $y_h$. 
This approach does not directly rely on the field-conjugate variable $\mathcal{Q}$ and its probability distribution function. 
Therefore, it is free from the explicit dependence of the mixing parameter and the histogram-visualization issue for the discrete data of $\mathcal{Q}$.  

We consider susceptibility, magnetization, specific heat, and compressibility as the thermodynamic quantities to be examined for our finite-size-scaling analysis. 
The susceptibility $\chi \equiv (L^d/T) (\langle |m|^2 \rangle - \langle |m| \rangle^2)$ and the magnetization $\langle |m| \rangle$ are estimated with the microcanonical magnetization by using Eq. (\ref{eq:mavg}). 
The specific heat $c \equiv (L^d/T^2) (\langle \epsilon^2 \rangle - \langle \epsilon \rangle^2)$ and the compressibility $\kappa_T \equiv (L^d/T) (\langle n^2 \rangle - \langle n \rangle^2)/\langle n \rangle^2$ are related to the fluctuations of the energy $E$ and the number $N$ of nonzero spins.
With the WL density of states being sampled with high accuracy, one can freely access these thermodynamic variables at any temperature and crystal field. 

\begin{figure}
\includegraphics[width=0.46\textwidth]{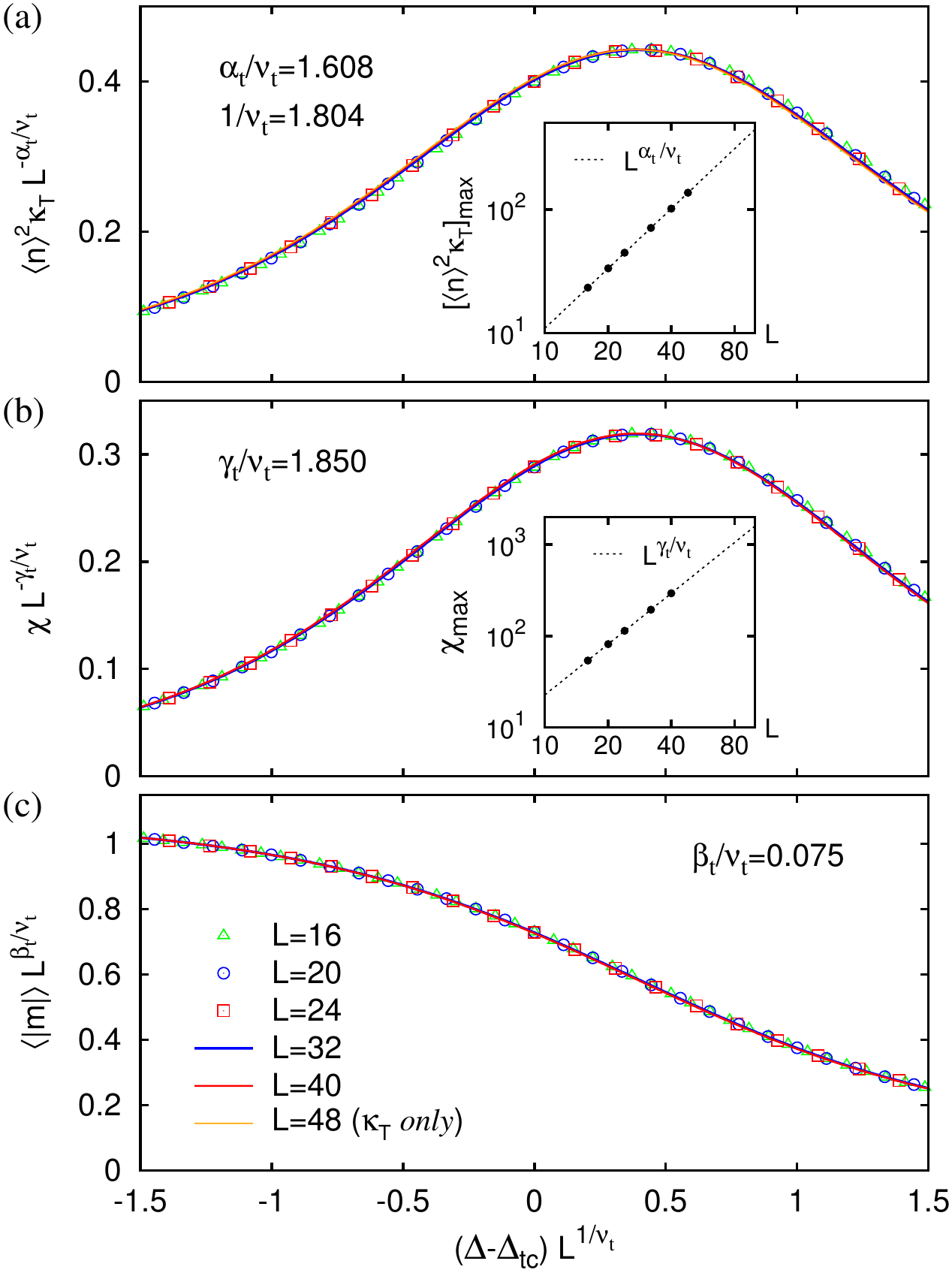}
\caption{\label{fig7}
(Color online) Tricritical behavior along the crystal field axis at the tricritical temperature. The finite-size-scaling analysis of (a) number fluctuations $\langle n \rangle^2 \kappa_T$, (b) susceptibility $\chi$, and (c) magnetization $\langle |m| \rangle$ is performed to determine the tricritical exponents. While the WL method guarantees high enough resolution to plot the data as continuous curves, the points in low resolution ($L\le 24$) are also given for visualization of finite-size scaling. The ratios $\alpha_t/\nu_t$ and $\gamma_t/\nu_t$ are determined from the power-law fits of the maxima of $\langle n \rangle^2 \kappa_T$ and $\chi$, respectively. Each scaling plot with the estimated exponent shows the excellent collapse of the data points falling onto a single curve. The tricritical eigenvalue exponents are deduced as $y_t=1.804$ and $y_h = 1.925$ from $\alpha_t/\nu_t$ and $\gamma_t/\nu_t$.    
}
\end{figure}

\begin{figure}
\includegraphics[width=0.46\textwidth]{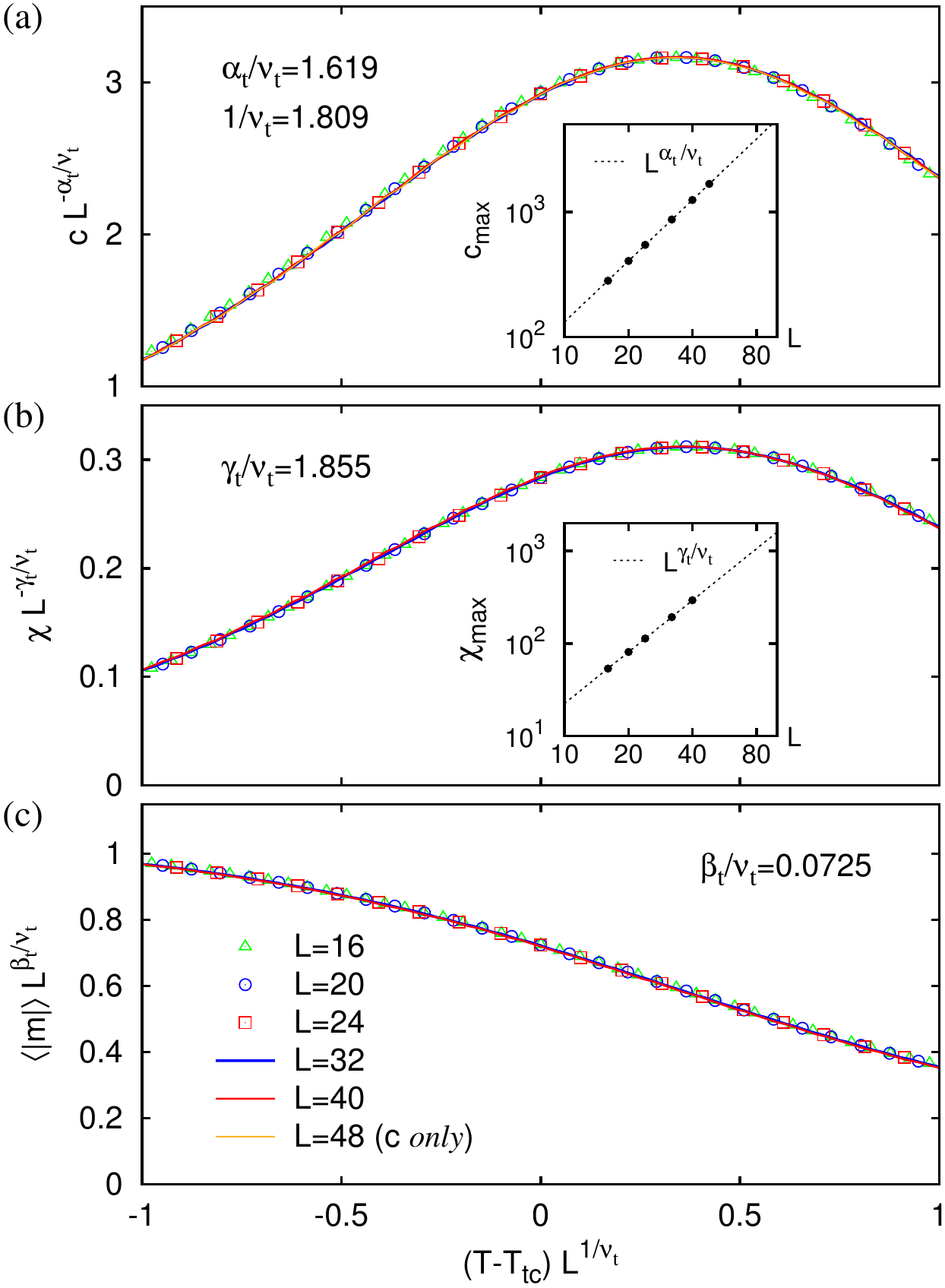}
\caption{\label{fig8}
(Color online) Tricritical behavior along the temperature axis. The fugacity is fixed at $\ln z \equiv \Delta/T=\Delta_{tc}/T_{tc}$. The finite-size scaling analysis of (a) specific heat $c$, (b) susceptibility $\chi$, and (c) magnetization $m$ is presented. The tricritical exponents are determined by the same procedures used in Fig.~\ref{fig7}. The resulting scaling plots show the excellent collapse of the data points falling onto a single curve, where the corresponding tricritical eigenvalue exponents are deduced to be $y_t=1.809$ and $y_h = 1.9275$ from $\alpha_t/\nu_t$ and $\gamma_t/\nu_t$.   
}
\end{figure}

Figures \ref{fig7} and \ref{fig8} show our finite-size-scaling analysis of the thermodynamic quantities for two different choices of an appropriate scaling axis. 
First, we choose to perform the finite-size scaling along the fugacity axis selected from the natural variables of the grand partition function. 
With the temperature fixed at $T=T_{tc}$, the scaling variable can be expressed as $x\equiv\Delta-\Delta_{tc}$. 
In this case, the relevant thermodynamic quantities are the number fluctuations, susceptibility, and magnetization, while the specific heat is discarded for our choice of the scaling test with fixed $T$. 
The corresponding scaling ansatz can be written as 
\begin{eqnarray}
\langle n \rangle^2 \kappa_T &=& L^{\alpha_t/\nu_t}\mathcal{N}^o(xL^{1/\nu_t}), \\
\chi &=& L^{\gamma_t/\nu_t}\chi^o(xL^{1/\nu_t}), \\
\langle |m| \rangle &=& L^{-\beta_t/\nu_t}\mathcal{M}^o(xL^{1/\nu_t}),
\end{eqnarray}  
where $\mathcal{N}^o$, $\chi^o$, and $\mathcal{M}^o$ are universal functions.
In comparison with Eq. (\ref{eq:P}), one can also obtain the relations between the conventional exponents, $\nu_t$, $\alpha_t$, $\beta_t$, and $\gamma_t$, through the tricritical eigenvalue exponents $y_t$ and $y_h$ as  
\begin{eqnarray}
\alpha_t/\nu_t &=& -d + 2 y_t. \\
-\beta_t/\nu_t &=& -d + y_h ,\\
\gamma_t/\nu_t &=& -d + 2 y_h.
\end{eqnarray} 
Provided the hyperscaling identity $\nu_t d = 2 - \alpha_t$, the thermal exponents are simply related as $y_t=1/\nu_t$.

The thermal exponent $y_t$ can be easily extracted from the maxima of $\langle n \rangle^2 \kappa_T$ which scales as $\langle n \rangle^2 \kappa_T \propto L^{\alpha_t/\nu_t}$. 
Figure~\ref{fig7}(a) shows the power-law fit of the maxima, providing the estimate of $\alpha_t/\nu_t=1.608$. 
This ratio of the exponents can be directly converted into the tricritical thermal exponent as $y_t=1/\nu_t=1.804$ which turns out to be very close to the exact conjecture $y_t=9/5$. 
The full finite-size-scaling ansatz for $\langle n \rangle^2 \kappa_T$ is also examined with the estimated exponents $\alpha_t/\nu_t=1.608$ and $1/\nu_t=1.804$, showing the excellent collapse of the data curves falling onto a single line, as shown in Fig.~\ref{fig7}(a).

We estimate the magnetic exponent $y_h$ through the similar analysis for the susceptibility of which maxima scales as $\chi \propto L^{\gamma_t/\nu_t}$. From the power-law fit shown in Fig.~\ref{fig7}(b), we find out $\gamma_t/\nu_t=1.850$, and this ratio is directly converted into the tricritical magnetic exponent $y_h=1.925$ which precisely agrees with the exact conjecture $y_h=77/40$. Figure~\ref{fig7}(b) shows the data perfectly falling onto a single curve in the test of the finite-size-scaling ansatz, confirming the accuracy of our estimate of the magnetic exponent. For the magnetization, while $\beta_t/\nu_t$ can be directly determined by the obtained $\gamma_t/\nu_t$ by using the scaling relations through $y_h$, we also examine the finite-size scaling ansatz of $\langle m \rangle$ for explicit confirmation, where we find the excellent collapse of the data curves falling onto a single line, as indicated in Fig.~\ref{fig7}(c)  

One the other hand, we perform another estimate of the tricritical exponents by choosing the $T$ axis for the similar finite-size-scaling analysis. 
The fugacity $z$ is now fixed at $\ln z\equiv\Delta/T=\Delta_{tc}/T_{tc}$, and thus the scaling variable is given as $x\equiv T-T_{tc}$. 
In this case, the relevant thermodynamic quantity for finite-size scaling includes the specific heat; namely, the energy fluctuations, instead of the number fluctuations. Although, the finite-size-scaling ansatz for the specific heat $c$ can be written similarly as 
\begin{equation}
c = L^{\alpha_t/\nu_t}\mathcal{C}^o(xL^{1/\nu_t}),
\end{equation}
where $\mathcal{C}^o$ is a universal function. 
The same scaling relation between $\alpha_t/\nu_t$ and $y_t$ holds for the specific heat as well.
By applying the same procedures as done for the earlier finite-size scaling in the $\Delta$-axis, here we estimate the tricritical exponents as $y_t=1.809$ and $y_h=1.9275$ on the $T$ axis, as shown in Fig.~\ref{fig8}. 
While the estimate of the tricritical exponents on the $T$ axis are slightly different from those estimated in the finite-size scaling on the $\Delta$ axis, both estimations are still in very good agreement with the exact conjectures, $y_t=9/5$ and $y_t=77/40$. 
The source of the discrepancy found between the two estimates may originate from the possibility that the error in locating the tricritical point propagates differently in our two choices of the scaling and fixed variables in the phenomenological finite-size-scaling analysis.

Finally, from the different forms of finite-size scaling that we have performed so far in this section, we can write the tricritical eigenvalue exponents of the BC model as
\begin{equation*}
y_t = 1.804(5), \quad y_g=0.80(1), \quad y_h=1.925(3),
\end{equation*} 
showing very good agreement with the exact conjectures, $y_t=9/5$, $y_g=4/5$, and $y_h=77/40$, respectively.
The estimated errors are mainly from the slight difference between the values observed in the different approaches of finite-size scaling. The comparison with the previous works using different numerical methods are also listed in Table \ref{table:exponent}.  

\section{Conclusions}
\label{sec:conclusions}

In conclusions, we have demonstrated the effectiveness of the Wang-Landau method in the finite-size-scaling analysis for tricritical behavior within the spin-$1$ Blume-Capel model in two dimensions. 
The significance of our results is two-fold.
First, we have constructed the detailed line of first-order transitions, completing the previously-less-explored area of the phase diagram at low temperatures, which is hardly accessible in conventional Monte Carlo simulations. 
In the area of large crystal fields very close to $\Delta=2$, we have found a double-peak structure in the specific heat where the Schottky-like anomaly is observed above the first-order-transition temperature.  
Second, through the various forms of the finite-size-scaling analysis, we have successfully estimated the tricritical point as $T_{tc} \simeq 0.6080$ and $\Delta_{tc}\simeq 1.9660$ and the tricritical exponents as $y_t = 1.804(5)$, $y_g=0.80(1)$, and $y_h=1.925(3)$. 
In particular, our high-resolution analysis of the phenomenological finite-size scaling takes a great advantage from the Wang-Landau methods, granting unrestricted access to the values of temperatures and crystal fields.     

The performance of the Wang-Landau method may depend on its practical limit in the system size which is still much smaller than those accessible in conventional methods. 
The large computational resource requirement is indeed one of the biggest obstacles that the Wang-Landau method should overcome to show its effectiveness in challenging problems of phase transitions. We have shown that, within the limit of our computational resources, the standard Wang-Landau algorithm now allows us to simulate the Blume-Capel model with sizes up to $48\times 48$ sites, which provide excellent finite-size scaling for the tricritical behavior. 
Our demonstration suggests that, with increasing computational power and potential support from more advanced techniques such as the recently suggested parallel algorithm for scalability \cite{Vogel2013,Vogel2014}, the Wang-Landau method may provide a promising tool of high-precision numerics for multicritical phenomena. 

\begin{acknowledgments}
This work was supported from Basic Science Research Program through the National Research Foundation of Korea funded by the Ministry of Science, ICT \& Future Planning (NRF-2014R1A1A1002682 D.H.K.,J.J., and J.L. and NRF-2013R1A1A2065043 for W.K.).  D.H.K., J.J., and J.L. also acknowledge support from the Top Brand Project of GIST. 
\end{acknowledgments}

\end{document}